# A quality-of-service mechanism for interconnection networks in system-on-chips

Wolf-Dietrich Weber*, Joe Chou, Ian Swarbrick, Drew Wingard
Sonics, Inc.® (www.sonicsinc.com).
* Google, Inc.®

## Abstract

*As Moore's Law continues to fuel the ability to build ever increasingly complex system-on-chips (SoCs), achieving performance goals is rising as a critical challenge to completing designs. In particular, the system interconnect must efficiently service a diverse set of data flows with widely ranging quality-of-service (QoS) requirements. However, the known solutions for off-chip interconnects such as large-scale networks are not necessarily applicable to the on-chip environment. Latency and memory constraints for on-chip interconnects are quite different from larger-scale interconnects. This paper introduces a novel on-chip interconnect arbitration scheme. We show how this scheme can be distributed across a chip for high-speed implementation. We compare the performance of the arbitration scheme with other known interconnect arbitration schemes. Existing schemes typically focus heavily on either low latency of service for some initiators, or alternatively on guaranteed bandwidth delivery for other initiators. Our scheme allows service latency on some initiators to be traded off smoothly against jitter bounds on other initiators, while still delivering bandwidth guarantees. This scheme is a subset of the QoS controls that are available in the SonicsMX™ (SMX) product.*

## 1 Introduction

Most SoCs consist of a number of intellectual property (IP) cores linked together via an on-chip "internal interconnect". Early SoC designs employed only a few cores, and could utilize well-established board-level bus schemes on the chip to satisfy the internal interconnect requirement. A switch from a tri-state driver based technology to a multiplex bus technology was adequate. But as the number of cores on chip grows, the complexity of the interaction between the IP cores increases exponentially, and a single computer bus is insufficient.

To satisfy internal interconnect requirements moving forward, multiple buses arranged as segments or hierarchies are the obvious next step. Full-blown networks on a chip [1,5, 6] are inevitable. These on-chip networks are required to meet the challenges imposed by the most advanced chip technologies [3, 4], and to enable the decoupling methodologies needed to address time-to-market pressures [1, 2].

Most of today's consumer devices such as set-top-boxes or mobile phones include some real-time complex data flows such as audio or video traffic, intermixed with more traditional processor-to-memory traffic. These real-time flows must be carefully managed to avoid compromising audio or video quality. Over-designing systems with separate resources such as buses and memory systems is one way to meet the requirements. But price pressure increasingly forces designers to share such critical resources as an external DRAM system. This results in several quality-of-service management challenges:

- Multiple initiators sending traffic to multiple shared targets
- Varying quality-of-service requirements for the different traffic flows, some with tight real-time or low-latency requirements
- Formation of a decentralized on-chip network with multiple points of arbitration

Little focus has been given on quality of service for internal interconnects [6 (Chapter 4)]. This paper proposes that quality-of-service management methodologies used in large scale interconnection networks [8, 9, 10,11], can be applied as a baseline for SoC internal interconnects. However, since on-chip interconnects are qualitatively very different, because they must deliver much lower latencies and typically cannot incorporate large amounts of intermediate storage, store and-forward of packets with software running smart prioritization algorithms are not possible. Therefore, the QoS mechanisms employed must be simple enough that they can be implemented in fast hardware without excessive storage.

The remainder of this paper is organized as follows:

Section 2. Operation of our Interconnect arbitration and how it fits with the quality-of-service model.

Section 3. Experiment Set Up Discussion (including system workloads)

Section 4. Presents and discusses the experiment results.

Section 5. Contrast of experiment results with related work

Section 6. Paper summary

## 2 Arbitration mechanisms

We focus on the arbitration of several initiators for a single shared target. Our solution breaks arbitration for the target into two parts, motivated by keeping complexity out of the interconnect core as much as possible (see Figure 1):

- arbitration in the core of the interconnect to deliver requests from different initiators to the target
- arbitration at the edge of the interconnect to enforce bandwidth allocations for different initiators

**Figure 1:** QoS mechanism distribution and network topology.

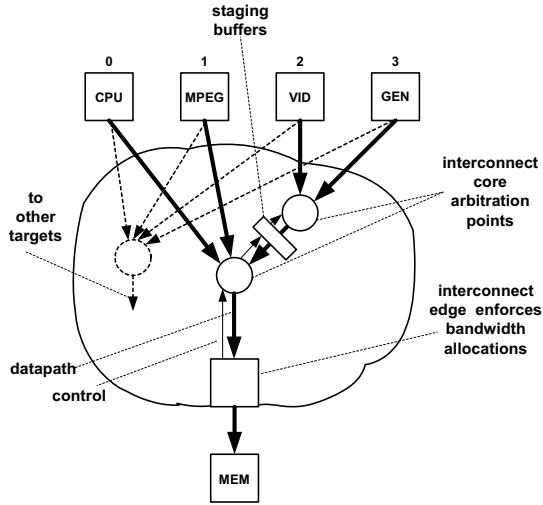

Since there can be multiple arbitration points in the interconnect, it is important that they each be efficient, and that their individual decisions combine to fulfil the QoS service model. By breaking the arbitration mechanisms into two parts, each can be optimized independently. The mechanisms in the core of the network must be simple and fast for minimum latency, while the mechanism at the edge of the network can be more sophisticated, and potentially target-specific.

Note that our description focuses on the request network here. The response network is similar but can generally have simpler arbitration mechanisms if no contention (i.e. only one active target per initiator) and no backpressure (i.e. initiator can always accept responses) is assumed.

Final arbitration at the interconnect edge is between different threads (or virtual channels [15]), each of which has been assigned to a quality-of-service level and has received a target bandwidth allocation. Initiators can either be given their own thread all the way to the target, or may share a thread with other initiators, depending on resource/performance trade-offs. In the core of the interconnect, threads each have their own buffering resources, and may therefore proceed independently of one another. The arbitration at the edge of the network sends information back into the interconnect (using sideband signals) to govern prioritization of threads within the interconnect. The network core and network edge arbitration combine to deliver the guarantees of the QoS model.

Threads leading to a target can be assigned one of the following quality-of-service levels (highest to lowest priority): priority threads, bandwidth threads, and best-effort threads. A given thread can be shared amongst several initiators, dedicated to a single initiator, or be one of several from a single initiator. Priority threads are optimized for low-latency service, bandwidth threads receive throughput guarantee within fixed jitter bounds, and best-effort threads receive service if and when bandwidth is left over by the other threads. Priority threads and bandwidth threads have a certain absolute target bandwidth allocation associated with them. As long as priority threads request service at a rate lower than their allocation, they receive absolute priority. This leads to low-latency service, which is why priority threads are typically used for initiators whose performance depends critically on the request service latency. When priority threads request service at a rate greater than their allocation, they are demoted to become best effort threads. This ensures that the QoS contracts of lower-priority threads can be honored. Bandwidth threads also receive an allocation and are serviced ahead of best-effort threads when within their allocation. Similar to priority threads, a bandwidth thread that requests service at a rate greater than its allocation is demoted to become a best-effort thread.

In the network core, arbitration resolves contention between requests from different initiators. A mechanism close to the target at the edge of the network keeps track of the recent bandwidth usage of each thread and instructs the interconnect core to demote a thread if necessary. Each of the two co-operating arbitration mechanisms is described in more detail below.

## 2.1 Epoch scheme

In the core of the network, an epoch scheme is used to govern arbitration. Each initiator sharing a target or set of targets sends a marker with every $n$th request, thus forming groups of requests or "epochs" of size $n$. The requests from different initiators that are contending for the same target are combined in such a way that no initiator may advance to the next epoch until all other initiators are ready to advance, or happen to have no requests to send. At the last arbitration point, a global epoch (including requests from all initiators) is formed. If all initiators are contending for access, each global epoch contains exactly one local epoch's worth of requests from each of the initiators. Since each initiator's epoch size can be set independently of all others, a non-uniform access scheme results. Moreover, if epoch markers for each initiator are inserted at the interconnect boundary, arbitration points within the core of the interconnect need not even know about the epoch sizes associated with each initiator. They merely combine requests so as to include one epoch's worth from each initiator. As shown in Figure 2, the arbitration point only allows requests from epoch 1 to proceed. When a request with an epoch marker shows up on one of the branches, that branch is removed from arbitration consideration, until all other branches either have a request with epoch marker waiting, or have no request pending. When this event occurs, the arbiter advances to the next epoch and considers requests from all branches again. Note that for requests from different initiators that are part of the same epoch, the epoch arbitration mechanism gives no preference to either request. A tie-breaker secondary arbitration mechanism is needed. We make use of a least-recently-serviced mechanism here, which tends to finely interleave requests from different initiators that are part of the same epoch.

**Figure 2:** Epoch scheme in operation.

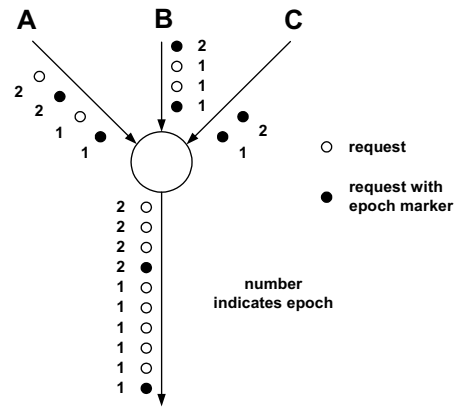

Since the epoch scheme is governed by markers associated with requests, it distributes well across arbitration trees, no matter how unbalanced their topology is. It is also robust through data width conversion, which may require aggregation or splitting of requests. Finally, it can easily handle frequency domain crossings, including asynchronous ones. These properties make it well-suited for use in high-speed on-chip networks for SOCs.

In the interconnect arbitration mechanism presented here, the epoch scheme is used for arbitration between different initiators sharing the same thread to a shared target. A version of the epoch scheme is also used to arbitrate amongst multiple threads at the same QoS level. For threads at different levels, a strict priority scheme is used, with the additional twist that priority and bandwidth threads can be demoted to bring them to the same level as best effort threads. The demotion is controlled from a bandwidth tracking mechanism at the edge of the network, close to the target core, which is described next.

## 2.2 Bandwidth allocation and enforcement

Bandwidth is allocated to different threads by the user and this allocation is enforced near the target core using a per-thread credit counter mechanism as shown in Figure 3. Each priority and bandwidth thread has a copy of this mechanism. The credit counter starts at 0 and is periodically incremented according to the bandwidth allocation. For example, if 25% of the target bandwidth has been allocated to a thread, it receives one credit every 4 cycles. When a request from a given thread is serviced, the corresponding credit counter is decremented. The counter thus keeps a moving window of bandwidth usage history over time. The counter has both an upper and lower saturation limit, set by the user. If a thread sends infrequent requests, its allocation count grows until it hits the positive limit, where it saturates. Similarly, if a thread sends requests more frequently than its allocation, the count decreases until it hits the negative limit. A thread that has a 0 count has been getting exactly as much service as its allocation allows.

**Figure 3:** Bandwidth allocation enforcement using counters.

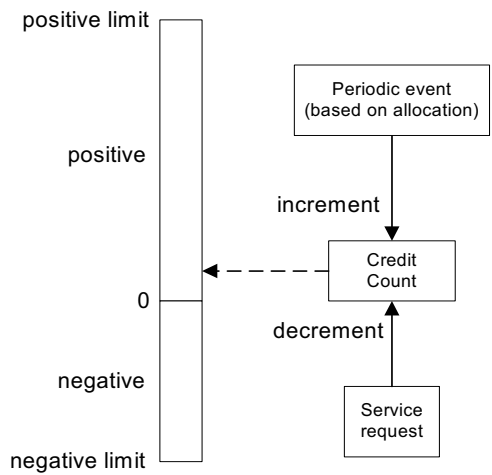

The credit counters are used to demote priority and bandwidth threads. If the count is negative, the corresponding thread is demoted.

The positive limit determines how much a given thread can request above its allocation, before it gets demoted. The positive limit must be at least large enough to account for the arrival jitter introduced by the request network. For threads with bursty arrival it is advantageous to set a higher positive limit, because it allows a larger burst to be serviced at once, without being smoothed out to the allocated bandwidth by demotion. It is often useful to set a larger positive limit for priority threads. However, there is a downside to a large positive limit: a large positive limit on priority threads can lead to large intervals of non-service for bandwidth threads (seen as increased jitter on the bandwidth threads).

The negative limit determines how long a thread's overusage of bandwidth will be remembered. If a thread is sending requests at a rate faster than its allocation, and that bandwidth is available (because other initiators are currently not sending requests), then the credit count becomes negative and is demoted. Once other initiators start requesting again, the demoted thread does not get service until its credit count becomes positive again. So a large negative limit can lead to larger service jitter after a period of service above the allocation.

## 2.3 Other arbitration schemes

Several other popular arbitration mechanisms are also used in the later experiments for comparison purposes.

One of the simplest mechanism for arbitration amongst several initiators is the *fixed priority* scheme. In this scheme, each initiator is assigned a position within a fixed priority order. Whenever a decision must be made between contending requests from different initiators, the request from the highest-priority initiator wins.

The fixed priority scheme is very unfair because low-priority initiators can be completely starved by high-priority initiators. A simple fair mechanism is the *round-robin* arbitration mechanism. In round-robin arbitration, there is also a fixed order of requestors, but the priority is adjusted dynamically after each successful arbitration to give the lowest priority to the previous winner. When all initiators are constantly requesting service, each initiator is serviced in turn.

The round-robin scheme gives every initiator equal access to the target. It may be desirable to have some initiators have a higher bandwidth of access to the target. The *time-division-multiplexed-access (TDMA)* scheme is one way of achieving this. Time is divided into equal-sized intervals, such as clock cycles, and each interval is assigned to a particular initiator. An uneven assignment of intervals to different initiators allows for different bandwidths to the target. Another characteristic of the TDMA mechanism is that time intervals are typically assigned to initiators in recurring periods, so that service is very regular. A downside of this strict assignment is that it does not allow an initiator to claim another initiator's interval (no preemption) which means that very-low latency service is typically not possible. Another downside of the TDMA mechanism is that it does not readily distribute to multiple arbitration points without including smoothing buffers between them to deal with misaligned TDMA wheels. This is especially true if there are data width conversion or clock domain crossings to accomplish.

The fixed weight scheme solves the round-robin equal allocation issue a different way: initiators are allowed to transfer more than one request when they are the highest-priority initiator, according to a set of fixed weights. The weights are set per arbitration point branch and are fixed in that arbitration point.

## 3 Experimental set-up

Simulations were performed at the cycle level. The models consisted of master models to drive a workload into the system, a slave model to service read and write requests, and an interconnect/arbitration model that contained the bulk of the different mechanisms under investigation. While bursts of requests are sourced by some initiators, arbitration takes place on a per-request basis. Also, the response network model is extremely simple, because no contention or backpressure are assumed.

The system set-up used in this paper is shown in Figure 4. We study four different initiators accessing a single shared target. For the early experiments, we generate artificial workloads to expose the characteristics of the different arbitration schemes. For the later experiments, we generate traffic that represents a more realistic system problem.

The data flows issued by the different initiators and the service requirements for these data flows are different for each initiator and represent a typical system problem with many conflicting performance goals. Initiator CPU represents a cached processor, with an 800 MHz internal core frequency, and assumed cpi of 1.0 resulting in a theoretical (no cache miss) upper bound of 800 MIPS. The performance of the processor (measured in MIPS) depends on the cache miss rate and the cache miss penalty. During the studies, we vary the cache miss rate to represent different application scenarios. The cache miss penalty is determined by the latency and bandwidth of the interconnect and target. Since the overall MIPS depends critically on the miss penalty when the cache miss rate is everything but negligibly small, a very important requirement for traffic from initiator CPU is that read requests be serviced with minimal latency.

The traffic generated by initiator CPU is a mix of 4-word burst reads and writes representing cache fills and writebacks. The characteristics are summarized in the table below.

**Table 1:** CPU traffic.

| Parameter | Value | Notes |
|---|---|---|
| burst size | 4 words | cache line size |
| RD/WR | 4:1 | |
| arrival | bursty, averaging a burst 4 cycles after last miss response | high miss rate (25% of instructions are loads/stores, 25% miss the cache, cpi = 1) |
| | bursty, averaging a burst 35 cycles after last miss response | low miss rate (25% of instructions are loads/stores, 2.85% miss the cache, cpi = 1) |

Initiators MPEG and VID represent stream-processing (audio, video, for example) initiators. These typically have some amount of buffering to deal with service interruptions, and are generally not very latency sensitive. The main service requirement for these data flows is that a minimum bandwidth must be sustained, and that bandwidth jitter must not exceed certain limits which would cause buffer overflow or underflow with a resultant loss of data.

The traffic generated by MPEG is somewhat bursty, with varying burst lengths, to represent the different alignment of memory blocks being read/written from memory.

**Table 2:** MPEG traffic.

| Parameter | Value | Notes |
|---|---|---|
| burst size | 1-8 words | varies dynamically |
| RD/WR | 2:1 | |
| arrival | bursty, 800 MB/s | |

We require the full maximum 800 MBytes/s bandwidth to properly service the MPEG traffic.

The traffic generated by VID represents reading a frame buffer and displaying it on a screen. Therefore it is all reads and very regular.

**Table 3:** VID traffic.

| Parameter | Value | Notes |
|---|---|---|
| burst size | 8 words | = smoothing buffer |
| RD/WR | all reads | |
| arrival | regular, 200 MB/s | |

We require 200 MBytes/s bandwidth to properly service the VID traffic.

Finally, initiator GEN represents some generic initiator that does not have high bandwidth needs and no specific service requirements (some non-system critical initiator).

**Table 4:** GEN traffic.

| Parameter | Value | Notes |
|---|---|---|
| burst size | 1-8 words | random |
| RD/WR | 1:1 | |
| arrival | bursty, 100 MB/s | |

All 4 initiators access a shared target MEM, which represents an on-chip SRAM core. For the MEM target we assume it has an 8-byte interface and runs at 200 MHz for a maximum bandwidth of 1.6 GBytes/s. It is modelled like an on-chip SRAM with full bandwidth capability and a latency of 1 cycle.

## 4 Experiments and results

This section describes the experiments and their results. Round-robin arbitration is not considered, since it is not especially interesting - the available resource is simply shared among all requestors.

### 4.1 System experiments

The experiments were run using the system workloads described in section 3. As shown in Figure 1, we have two arbitration points. Initiators VID and GEN arbitrate at one point, and the result is combined with requests from CPU and MPEG at a second arbitration point. We study priority, TDMA, and epoch-based arbitration with dynamic demotion in the two arbitration points.

When allowing multiple initiators to share a target, there needs to be some way to assign a portion of the service capability of the tar-get to the different initiators. For our system scenario, we require the CPU initiator to be serviced with minimum latency, while the MPEG and VID initiators need a certain bandwidth within the given jitter limits.

A simple way to guarantee low-latency access to the CPU is to use a fixed-priority scheme and give the highest priority to the CPU. Other initiators are assigned lower priorities in decreasing order of bandwidth, i.e. MPEG is next, followed by VID and GEN.

The results for a low CPU miss rate are shown in Figure 4. Each initiator gets its required bandwidth and the CPU achieves 678 MIPS.

**Figure 4:** System results, priority scheme (low miss scenario)

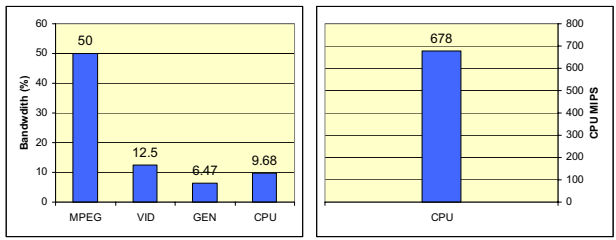

However, when the CPU miss rate increases to the high miss rate, as shown in Figure 5, it ends up stealing too much bandwidth from the other initiators, and the VID core does not get its required bandwidth. This would manifest itself as screen flicker, which is very undesirable.

**Figure 5:** System results, priority scheme (high miss scenario)

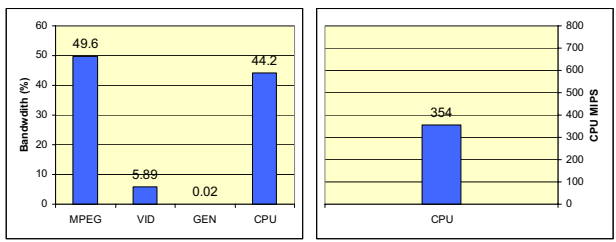

At the other end of the spectrum, we can use a TDMA scheme to assure that the MPEG and VID initiators get their required bandwidth within tight jitter bounds. We allocate every second cycle to the MPEG initiator, every 4th to the CPU, and one each out of 8 to the VID and GEN initiators.

The results are shown in Figures 6 and 7. While the MPEG and VID initiators now receive the bandwidth they require, and the jitter is naturally very low in this scheme, the CPU latency has gone up dramatically. For the low miss rate we now see only 524 MIPS, which is only 77% of what the priority scheme achieves. For the high miss rate, the degradation is even worse, as the CPU dips below 167 MIPS.

**Figure 6:** System results, TDMA scheme (low miss scenario)

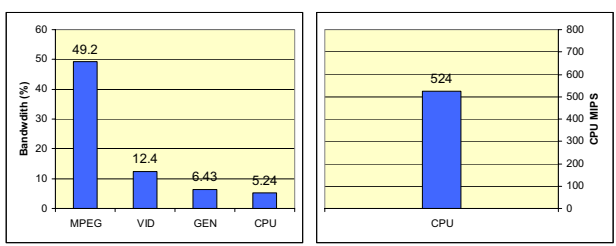

**Figure 7:** System results, TDMA scheme (high miss scenario)

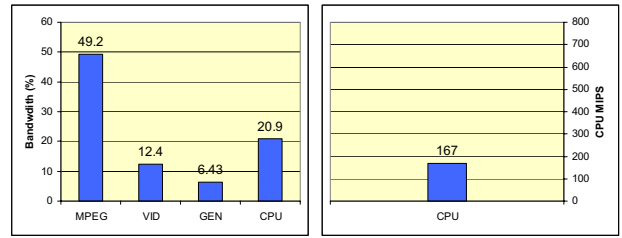

Our arbitration scheme with bandwidth allocation achieves the best of both worlds. We have allocated 800 MBytes/s to MPEG, 240 MBytes/s to VID, and the rest to the CPU. With low CPU misses (Figure 8) we see that the MPEG and VID initiators receive their required service, while the CPU gets to the same 678 MIPS that the priority scheme achieved. When the CPU miss rate increased to the high setting (Figure 9), MPEG and VID remain protected (as in TDMA) but the CPU achieves a MIPS value of 280, which is 68% higher than the TDMA scheme.

**Figure 8:** System results, QoS scheme (low miss scenario)

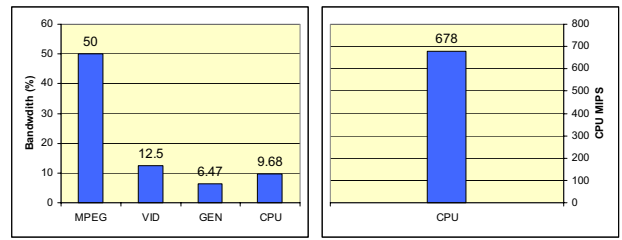

**Figure 9:** System results, QoS scheme (high miss scenario)

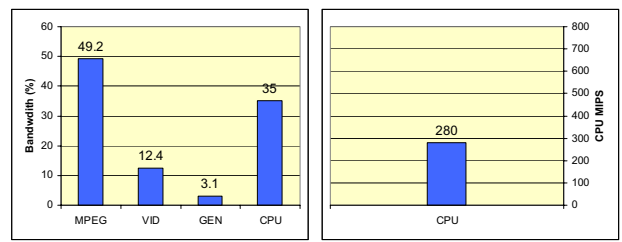

The CPU thread has been optimized for latency. When within its allocation, the CPU always sees the minimum request service latency of 1 cycle. Once above its allocation (as happens frequently in the high miss scenario), service drops back to the allocated bandwidth and queueing latencies are much higher. The MPEG and VID initiators are allocated their required bandwidth.

## 5 Related work

There has been much research into providing quality-of-service guarantees in high-speed networks. Various service models and traffic models for packet-switching networks that provide (deterministically) guaranteed service, best-effort service, predicted service, and statistically guaranteed service have been described [8, 9, 10, 11]. For on-chip interconnects, latency requirements are typically much tighter, making solutions that require a set-up phase for a connection impractical. Complex algorithms that require much intermediate storage also do not readily apply to on-chip, all hardware implementations. Fortunately, the on-chip problem is typically more bounded in terms of interconnect size, distance in clock cycles, and initiator traffic behavior, so a mechanism that provides deterministic (rather than statistical) guarantees is feasible.

Work focussed on quality-of-service specifically for on-chip interconnects is much sparser and more recent. One of the major challenges in implementing a shared on-chip interconnect (without overdesigning) is to provide service guarantees to some initiators while at the same time keeping resource utilization high. We are aware of three other efforts to give both some form of quality-of-service guarantee while at the same time allowing for best-effort traffic to keep utilization high. The SiliconBackplane mechanism [16] makes use of a TDMA wheel for guaranteed throughput, and allows unused slots to be filled by "best effort" traffic. It does not, however, provide any optimizations for low-latency service. The AETHEREAL NOS [12, 13] emphasizes the need for guaranteed services in systems-on-chips and provides a router implementation that combines best-effort with guaranteed services. From this standpoint, we are very much aligned with the thrust of this work. The major differences with our work is that we not only provide bandwidth guarantees, but also offer a service class optimized for low-latency. The LOTTERYBUS [14] addresses the long latency issues of traditional TDMA schemes by implementing a statistical TDMA to enforce bandwidth allocations. This helps somewhat, but we go further by giving priority to initiators with requirements for low-latency service, rather than just a statistical chance amongst contending initiators. The only way to achieve low latency in LOTTERYBUS is to overallocate bandwidth to the corresponding initiator, making the rest of the system vulnerable to bandwidth spurts from that initiator. Note that in LOTTERYBUS statistical rather than absolute guarantees are given, which means system designers must still account for the (small) chance that a bandwidth allocation fails. Finally, there seems to be no obvious way to extend this scheme to more than a single arbitration point.

## 6 Summary

This paper has outlined a simple quality-of-service scheme for on-chip interconnects. It offers service guarantees to each initiator, regardless of the other initiators' offered traffic load, which greatly simplifies the system performance optimization problem. Three levels of quality of service are available for each initiator: priority (optimized for low-latency up to a maximal throughput), bandwidth (offering a guaranteed throughput) and best-effort (no service guarantees). While some existing schemes can offer bandwidth guarantees, we offer the ability to provide both low-latency service and bandwidth guarantees. Initiators that require low latency will always experience the lowest possible latency *except* in the situation where servicing them will violate the bandwidth guarantees of other initiators. This feature is critical for sharing resources between latency-sensitive dataflows from initiators such as general-purpose processors and real-time dataflows such as video traffic, which is an increasingly common scenario in SOCs. The functionality of the arbitration mechanisms distributes well over non-uniform interconnects with multiple arbitration points, and naturally handles such bandwidth discontinuities as data width conversion and clock domain crossings. Division into network core and network edge functions allows for very high-speed implementations.